\documentclass[sigconf]{acmart}
\AtBeginDocument{%
  }

\setcopyright{acmlicensed}
\copyrightyear{2018}
\acmYear{2018}
\acmDOI{XXXXXXX.XXXXXXX}
\acmConference[Conference acronym 'XX]{Make sure to enter the correct
  conference title from your rights confirmation email}{June 03--05,
  2018}{Woodstock, NY}
  
\acmISBN{978-1-4503-XXXX-X/2018/06}
\setcopyright{none}
\renewcommand\footnotetextcopyrightpermission[1]{}
\usepackage{caption}
\usepackage{subcaption}
\usepackage{multirow}
\usepackage{enumitem}
\usepackage{makecell}

\begin{document}

\title{Context-Aware Disentanglement for Cross-Domain Sequential Recommendation: A Causal View}

\author{Xingzi Wang}
\email{wangxz@stu.sufe.edu.cn}
\orcid{0009-0009-7623-4704}
\affiliation{%
  \institution{School of Computing and Artificial Intelligence, Shanghai University of Finance and Economics}
  \city{Shanghai}
  \country{China}
}

\author{Qingtian Bian}
\email{bian0027@e.ntu.edu.sg}
\affiliation{%
  \institution{College of Computing and Data Science, Nanyang Technological University}
  \country{Singapore}}

\author{Hui Fang}
\email{fang.hui@mail.shufe.edu.cn}
\affiliation{%
  \institution{School of Computing and Artificial Intelligence, Shanghai University of Finance and Economics}
  \city{Shanghai}
  \country{China}
}

\renewcommand{\shortauthors}{Wang et al.}

\vspace{-6mm}
\begin{abstract}
Cross-Domain Sequential Recommendation (CDSR) aims to enhance recommendation quality by transferring knowledge across domains, offering effective solutions to data sparsity and cold-start issues. However, existing methods face three major limitations: (1) they overlook varying contexts in user interaction sequences, resulting in spurious correlations that obscure the true causal relationships driving user preferences; (2) the learning of domain-shared and domain-specific preferences is hindered by gradient conflicts between domains, leading to a seesaw effect where performance in one domain improves at the expense of the other; (3) most methods rely on the unrealistic assumption of substantial user overlap across domains. To address these issues, we propose CoDiS, a context-aware disentanglement framework grounded in a causal view to accurately disentangle domain-shared and domain-specific preferences. Specifically, Our approach includes a variational context adjustment method to reduce confounding effects of contexts, expert isolation and selection strategies to resolve gradient conflict, and a variational adversarial disentangling module for the thorough disentanglement of domain-shared and domain-specific representations. Extensive experiments on three real-world datasets demonstrate that CoDiS consistently outperforms state-of-the-art CDSR baselines with statistical significance.
\vspace{-1mm}
\end{abstract}




\maketitle
\vspace{-1mm}
\section{Introduction}

Sequential recommendation, crucial for platforms like Amazon and
YouTube, models user preferences through interaction sequences. However, traditional single-domain approaches often encounter persistent challenges such as data sparsity and cold-start issues, which impact recommendation quality and user satisfaction. To overcome these problems, Cross-Domain Sequential Recommendation (CDSR) has emerged as a promising solution that facilitates knowledge transfer across interconnected domains. This makes CDSR particularly valuable in scenarios like entering new markets and leveraging multi-platform recommendation services.

Recent advances in CDSR mainly focus on learning two types of preference representations: domain-specific preferences from individual domain sequences and domain-shared preferences from combined cross-domain sequences~\cite{2023dream,cao2022c2dsr}. These methods often use advanced representation transfer or alignment techniques to improve recommendation performance in the target domain. Despite significant progress, there remain three major limitations that constrain their effectiveness in real-world applications.

\begin{figure*}[t]
    \centering
    \setlength{\abovecaptionskip}{-0.5mm}
    \includegraphics[width=\linewidth]{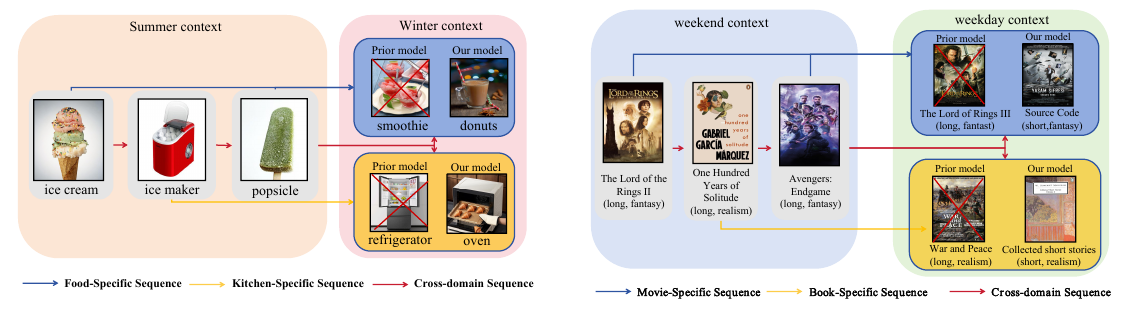}
    \caption{CDSR comparison of prior models and our model examples under varying contexts. 
    (a) Prior model would misinterpret spurious correlation as cross-domain preferences.
    (b) Prior model would spuriously correlate domain-specific preference.}
    \label{fig:varying_context_examples}
    \vspace{-3mm}
\end{figure*}

First, existing CDSR methods often overlook the critical influence of varying context in user interaction sequences. These contexts, such as seasonal trends or shifts in popularity over time~\cite{gong2021aesthetics}, can significantly influence how user preferences are expressed. In real-world CDSR scenarios, contextual influence is highly complex, leading current methods to potentially learn spurious correlations and produce inaccurate predictions when modeling domain-shared and specific preferences. As shown in \autoref{fig:varying_context_examples}, two key scenarios exemplify this issue: 1) Context influencing cross-domain shared preferences. In a food-kitchen scenario, users maintain a stable domain-shared preference for sweet tastes, but its expression varies seasonally (ice cream in summer, hot cocoa in winter). Existing models often regard these seasonal factors as truly shared preferences and continue recommending cold items, ignoring the underlying context.
2) Context spuriously correlates domain-specific preferences. In a movie-book scenario, a user may have a domain-specific preference for fantasy movies and realistic books separately, yet weekend contexts lead to longer content consumption. Models may misattribute this common tendency to the shared preference, leading to persistent recommendations of long content in both domains, even on weekdays. These issues highlight a major limitation in existing CDSR methods: they cannot reliably distinguish genuine user preferences (shared or specific) from spurious correlations induced by contexts. This results in suboptimal recommendations.

Second, prior models face challenges in disentangling domain-shared and domain-specific preferences. As highlighted, the spurious correlations induced by context further hinder this process, resulting in mixed and blurred representations. Moreover, the seesaw effect, a common issue in multi-task learning, where improving one domain's performance often harms the other due to gradient conflicts. This effect prevents accurate discovery of domain-shared preferences and causes negative transfer between domains.

Third, current CDSR methods rely heavily on bridging mechanisms that assume substantial user overlap between domains to enable knowledge transfer. This assumption often fails in practical scenarios, where overlapping users across domains are rare, limiting the applicability of existing models in such conditions.

To overcome these obstacles, we propose CoDiS, a robust framework for disentangling domain-specific and shared preferences while mitigating context confounding and negative transfer. CoDiS introduces a variational context adjustment mechanism that approximates contextual variables and performs backdoor adjustment during representation learning. Specifically, we adopt context-aware Mixture-of-Experts (MoE) encoders with expert isolation and selective routing to dynamically assign shared or specific experts based on context while avoiding gradient conflicts. Furthermore, CoDiS incorporates a variational adversarial disentanglement module, where a domain discriminator coupled with a Gradient Reversal Layer (GRL) encourages domain-shared and domain-specific representations decoupling. Importantly, CoDiS remains effective even under non-overlap conditions by capturing causal preferences, eliminating reliance on explicit cross-domain user alignment, making it highly applicable to real-world cross-domain scenarios.

Extensive experiments conducted on three real-world datasets demonstrate that CoDiS consistently outperforms state-of-the-art (SOTA) CDSR models across all metrics. Furthermore, its ability of knowledge transfer under conditions with sparse or non-existent user overlap highlights its robustness and practical applicability in diverse real-world scenarios. 

The main contributions of this paper are summarized as follows:
\begin{itemize}[topsep=0pt,itemsep=0pt,parsep=0pt,leftmargin=4mm]
    \item CoDiS is the first work to perform disentangled representation learning from a causal perspective in CDSR. It employs a \textbf{variational context adjustment} mechanism to eliminate the confounding effects of contextual information, thereby ensuring a more accurate modeling of user preferences.
    \item CoDiS introduces \textbf{experts isolation and selection} strategies alongside \textbf{variational adversarial disentanglement} to mitigate gradient conflicts, improving the independency of shared and domain-specific preferences.
    \item CoDiS removes the dependency on overlapping users, enabling successful knowledge transfer even under conditions with sparse user overlap.
    \item Extensive experiments confirm that CoDiS significantly outperforms previous SOTA approaches on three real-world datasets across multiple metrics.
\end{itemize}

\section{Related Work}
\subsection{Cross-Domain Recommendation (CDR)}
Cross-Domain Recommendation (CDR) addresses data sparsity by leveraging transfer learning techniques. Common approaches include domain alignment, which aligns user or item representations across different domains~\cite{ma2024triple,wang2021low,zhao2023cross}, and domain adaptation, which transfers knowledge from source domains to enhance target domains~\cite{hu2018conet,li2020ddtcdr}. However, indiscriminate transfer risks negative transfer by leaking domain-specific biases. This has motivated the development of disentanglement-based CDR methods~\cite{zhang2023multi,cao2022disencdr,guo2023disentangled,wang2025enhancing}. Recently, causal inference has been explored to learn invariant user preferences across domains~\cite{menglin2024c2dr,du2024identifiability,zhu2025causal,zhang2024transferring}. Nevertheless, these methods are primarily designed for non-sequential data and struggle to handle sequential recommendation due to dynamically evolving contexts and complex temporal dependencies.

\vspace{-5mm}
\subsection{Cross-Domain Sequential Recommendation (CDSR)}
CDSR extends CDR into sequential settings by leveraging dynamic user behaviors across domains. Early works ~\cite{ma2019pi,sun2023psjnet,chen2019dagcn} focus on transferring sequential knowledge across domains with shared account, which primarily focus on domain-level knowledge transfer without explicitly disentangling cross-domain and domain-specific information and completely dependent on user overlap. More recent approaches~\cite{2023dream,cao2022c2dsr,xu2025multi} introduce disentangled representations to separate domain-specific and shared preferences. However, they often suffer from inter-domain gradient conflicts and spurious correlation, and remain dependent on user-overlapping sequences. Recently, some studies have explored CDSR under user non-overlap settings~\cite{xu2024rethinking,lin2024mixed,xu2025heterogeneous}. However, their effectiveness relies heavily on strong item overlap or consistent latent group structures, which often fail to capture true causal preferences and may instead model spurious correlations.

\vspace{-1mm}
\section{Preliminaries}
\subsection{Problem Formulation}
In this study, we focus on the dual-target CDSR task, involving two distinct domains denoted as $A$ and $B$. We denote $\mathcal{U} = \{1, 2, \cdots, |\mathcal{U}|\}$ as the set of users and $\mathcal{I}^{A}$, $\mathcal{I}^{B}$ as the item spaces for domain $A$ and domain $B$, respectively. For each user $u \in \mathcal{U}$, we represent their interaction sequences in the two domains as $S^{A}=(i^{A}_{1}, i^{A}_{2}, i^{A}_{3}, ..., i^{A}_{|S^A|})$ with $i^{A} \in \mathcal{I}^{A}$, and $S^{B}=(i^{B}_{1}, i^{B}_{2}, i^{B}_{3}, ..., i^{B}_{|S^B|})$ with $i^{B} \in \mathcal{I}^{B}$. The goal of dual-target CDSR is to predict the next item $Y^A=i^{A}_{n+1}$ and $Y^B=i^{B}_{n+1}$ that the user may interact with in both domains. As mentioned before, the data distribution is normally affected by time-dependent external factors, i.e., context $C$. The task can be formulated as:

\noindent\textbf{Input}: One user’s domain-specific sequences $S^{A}$ and $S^{B}$.

\noindent\textbf{Output}:  The estimated probability of this user’s next interaction items in both domains:
\vspace{-1mm}
\begin{equation}
\arg\max_{Y^A \in \mathcal{I}^{A}} P(Y^A \mid S^{A}, S^{B}, C), \quad
\arg\max_{Y^B \in \mathcal{I}^{B}} P(Y^B \mid S^{A}, S^{B}, C).
\end{equation}
\vspace{-5mm}

\begin{figure}[htbp]
    \centering
    \includegraphics[width=\linewidth]{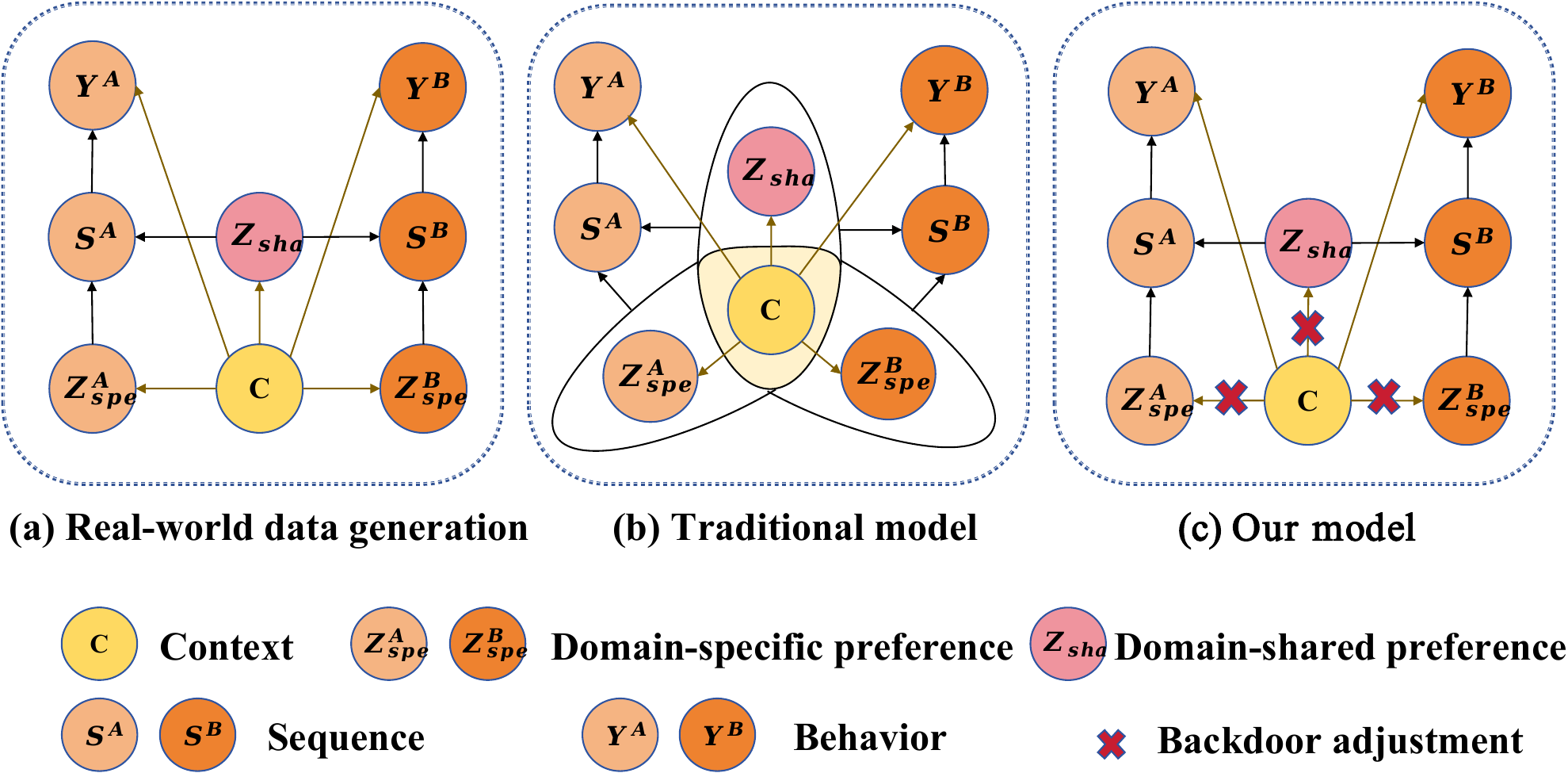}
    \setlength{\abovecaptionskip}{-1mm}
    \caption{
    A comparison of real-world data generation, the traditional model, and our model in CDSR scenario.
  }
  \label{fig:causal_view}
  \vspace{-6mm}
\end{figure}

\subsection{Causal Data Generation View of CDSR}

\subsubsection{\textbf{Confounding Effects of Contexts}}
In CDSR scenarios, the true data-generating process involves contextual factors (\(C\)) acting as confounders that influence shared preferences (\(Z_{\text{sha}}\)), domain-specific preferences (\(Z^A_{\text{spe}}, Z^B_{\text{spe}}\)), and observed behaviors (\(S^A, S^B\) and \(Y^A, Y^B\)), as shown in~\autoref{fig:causal_view}a. This introduces two major issues in traditional models (\autoref{fig:causal_view}b): 
First, due to \(C\) simultaneously influencing \(Z\) and \(Y\), the backdoor path \(Y \leftarrow C \rightarrow Z\) leads models to capture false dependencies, misinterpreting context-driven patterns as genuine user preferences. 
Second, the influence of \(C\) entangles shared preferences (\(Z_{\text{sha}}\)) and domain-specific preferences (\(Z^A_{\text{spe}}, Z^B_{\text{spe}}\)) with contextual factors, making them non-independent. This dependency limits the model's ability to fully disentangle these preferences, as they are always mixed with the effects of \(C\), reducing the clarity and effectiveness of the learned representations.

\subsubsection{\textbf{Our Causal Intervention}}
An ideal way to compute $P(\text{do}(Z))$ is to carry out randomized controlled trial (RCT)~\cite{pearl2016causal} by recollecting data from large-scale randomized samples under any possible
context, which is infeasible. Fortunately, there exists a statistical estimation of $P(\text{do}(Z))$ by leveraging backdoor
adjustment. To mitigate these biases, we adopt do-calculus to cut the backdoor paths, ensuring unbiased inference of preferences:

\vspace{-2mm}
\begin{equation}
\begin{aligned}
P(\text{do}(Z)) &= \sum_{i=1}^{|C|} P(Z \mid C = c_i) P(C = c_i).
\end{aligned}
\label{eq:backdoor_adjustment}
\end{equation}

Direct optimization of \(P(\text{do}(Z))\) is challenging due to the unobservability of \(C\) and the unknown prior \(P(C)\). Like in ~\cite{yang2022towards}, we approximate \(C\) using the variational posterior \(Q(C \mid S)\) and derive the Evidence Lower Bound (ELBO) as: 

\vspace{-1mm}
\begin{equation}
\begin{aligned}
\log P_\theta(Y \mid S,\text{do}(Z)) \geq &\mathbb{E}_{c \sim Q(C \mid S)} \big[ \log P_\theta(Y \mid S, Z, C = c) \big] \\
&- D_{\text{KL}}(Q(C \mid S) \| P(C)),
\end{aligned}
\label{eq:elbo}
\end{equation}
\vspace{-1mm}

\noindent where the last step is given by Jensen’s Inequality and the equality holds if and only if \(Q(C \mid S)\)
exactly fits the true posterior $P(C \mid S, Y )$, which suggests it successfully uncovers the latent context from observed data. To estimate \(P(C)\), we adopt the approach proposed in ~\cite{yang2022towards} by using a mixture of pseudo variational posteriors:
\vspace{-2mm}
\begin{equation}
\hat{P}(C) = \frac{1}{V} \sum_{j=1}^{V} Q(C \mid S = S_j'),
\end{equation}
\vspace{-1mm}

\noindent where \(V \ll N\) and \(S_j'\) is a randomly generated pseudo event sequence. This method ensures a flexible and data-driven estimation of the prior while reducing computational costs. Our approach eliminates confounding effects and achieves unbiased learning of shared and domain-specific preferences.
\vspace{-1mm}

\section{Methodology}
\begin{figure*}[t]
  \centering
    \setlength{\abovecaptionskip}{-1mm}
    \includegraphics[width=\linewidth]{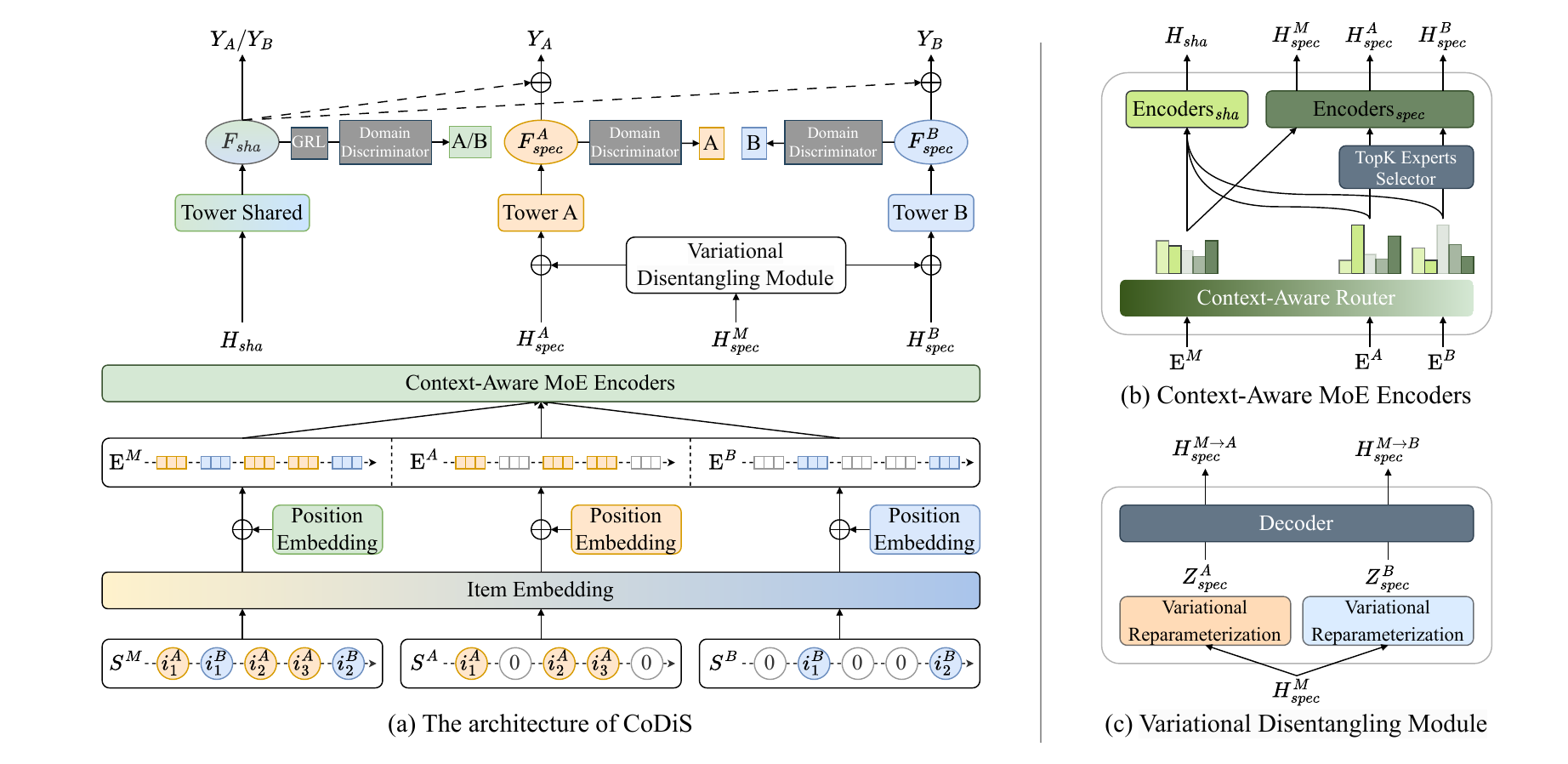}
  \caption{(a) The architecture of CoDiS. (b) The structure of context-aware MoE Encoders. (c) The structure of the variational disentangling module.}
\label{fig:overview of CoDiS}
\vspace{-3mm}
\end{figure*}

In this section, we propose CoDiS as shown in \autoref{fig:overview of CoDiS}. The
introduction of CoDiS comprises five subsections: (1) Sequence Formulation and Embedding; (2) Context-Aware MoE Encoders; (3) Variational Disentangled Module; (4) Adversarial Disentangling Module; (5) Model Training.

\vspace{-1mm}
\subsection{Sequence Formulation and Embedding}
Given a user’s raw training sequences $S^A$ and $S^B$, we first construct the mixed sequence $S^M$ by merging $S^A$ and $S^B$ in chronological order based on timestamps ($S^M$ is identical to $S^A$ or $S^B$ if the user has interactions in only one domain). Then, we pad all the sequences to the same maximum length T. Each item is then embedded into a learnable vector of dimension $h$. The item embedding matrix is defined as $\mathbf{W}_\mathrm{item} \in \mathbb{R}^{L \times h}$, where $L$ is the total number of unique items. Similarly, the position embedding matrix is set as $\mathbf{W}_\mathrm{pos} \in \mathbb{R}^{T \times h}$. Both embedding matrices are shared across all sequences. For each sequence, its representation is obtained by summing the corresponding item and position embeddings, followed by a dropout layer for regularization. The resulting sequence embeddings are denoted as $\mathbf{E}^A = [e_1^A, \mathtt{[PAD]}, \dots, \mathtt{[PAD]}, e_T^A]$
, $\mathbf{E}^B = [\mathtt{[PAD]}, e_2^B, ..., e_{T-1}^B, \mathtt{[PAD]}]$, and $\mathbf{E}^M = [e_1^A, e_2^B, ...,e_{T-1}^B, e_T^A]$, respectively. $\mathtt{[PAD]}$ denotes a padding token.

\vspace{-1mm}
\subsection{Context-Aware MoE Encoders}
This module comprises a router to infer context for context-specific experts to capture context-invariant sequential patterns, and expert isolation to prevent gradient conflicts between domains, addressing the seesaw effect and context variation.
\subsubsection{\textbf{Context-Specific Expert}}
To effectively encode the input sequences while accounting for the influence of the context $C$ on recommendations, we design an expert $\Phi(\cdot)$ that is conditioned on a specific context. Motivated by the success of the self-attention encoder in sequential modeling~\cite{kang2018self}, we employ it as the backbone for $\Phi(\cdot)$. 

We assume 
$N$ distinct contexts in the cross-domain space, i.e., $\mathcal{C} = \{\mathbf{c}_n\}_{n=1}^N$, each represented by an $N$-dimensional one-hot vector $\mathbf{c}_n \in \{0, 1\}^N$, with the $n$-th entry set to 1 and all others to 0.

In the embedding layer, a user's interaction history is embedded. Thus, we construct a set of context-specific experts $\{\Phi_n(:; \theta_n)\}_{n=1}^N$ to model context-aware preferences, where $\Theta = \{\theta_n\}_{n=1}^N$ is the collection of expert parameters. The context-specific representations are given by:

\vspace{-4mm}

\begin{equation}
\begin{aligned}
    H_n^A = \Phi_n(\mathbf{E}^A; \theta_n),\quad H_n^B = \Phi_n(\mathbf{E}^B; \theta_n), \quad H_n^M = \Phi_n(\mathbf{E}^M; \theta_n).
\end{aligned}
\end{equation}

\subsubsection{\textbf{Context-aware Router: Dynamic Expert Selection}}
We next introduce a Context-aware Router $\Psi(\cdot)$ to parameterize the variational posterior $Q(C|S)$. The goal is to dynamically infer the context assignment $c_{(t)}$ given the sequence. At each time step, $\Psi(\cdot)$ takes the sequence representation as input and outputs a probability vector $\mathbf{q}_t \in [0,1]^N$, where the $n$-th entry indicates the probability of the corresponding context $c_n$, i.e., $\mathbf{q}_t = \Psi(e_t; \Omega)$.

To achieve this, we parameterize each context $c_n$ with a learnable embedding matrix $\mathbf{H}_c \in \mathbb{R}^{N \times h'}$ where $h' = h(h+1)$. The embedding for the n-th context is given by $\mathbf{w}_n = \mathbf{c}_n^\top \mathbf{H}_c$. We then split each $\mathbf{w}_n$ into several parameters:

\vspace{-4mm}
\begin{equation}
    \mathbf{W}_n = \mathbf{w}_n[:h^2].\operatorname{reshape}(h, h), \quad \mathbf{a}_n = \mathbf{w}_n[h^2 : h(h+1)],
\end{equation}

\noindent where $\mathbf{W}_n \in \mathbb{R}^{h \times h}$, $\mathbf{a}_n \in \mathbb{R}^{h}$. The attribution score $s_{tn}$ that measures how likely a sequence up to time step $t$ belongs to context $\mathbf{c}_n$ can be calculated via:

\vspace{-5mm}
\begin{equation}
    s_{tn} = \langle \mathbf{a}_n, \ \mathrm{Swish}(\mathbf{W}_n \mathbf{e}_t) \rangle,
\end{equation}
where $\mathbf{e}_t \in \mathbb{R}^h$ is the embedding of the sequence up to step $t$, and $\langle \cdot, \cdot \rangle$ indicates the inner product operation. The Swish function is defined as $\mathrm{Swish}(x) = x \cdot \mathrm{Sigmoid}(x)$ \cite{touvron2023llama}.

\subsubsection{\textbf{Shared Experts Isolation and Specific Expert Selection}}
As mentioned, there are $N$ types of context-specific experts. Among them, the first $R$ experts are shared across domains, while the remaining $N-R$ are domain-specific. For each input sequence embedding $\mathbf{E}^X$ where $X \in \{A,B,M\}$, the gating mechanism computes expert weights through masked selection:
\vspace{-1mm}
\begin{equation}
\mathbf{q}_t^{X} = \mathrm{Softmax}\left(
\left[ s_{t1}^X, \dots, s_{tN}^X \right] \odot \mathbf{m}^X
\right),
\end{equation}
\vspace{-3mm}

\noindent where the mask matrix $\mathbf{m}^X \in \{0,1\}^N$ enforces expert selection rules:
\vspace{-1mm}
\begin{equation}
\mathbf{m}_n^X = \begin{cases} 
1, & \text{if } n \leq R \text{ or } n \in \mathrm{TopK}_k\left(\{s_{tj}^X\}_{j=R+1}^N\right) \\ 
0, & \text{otherwise}
\end{cases},\, X \in \{A,B\}.
\end{equation}
For the mixed sequence M, all experts are activated with $m^{M} \equiv 1$. The shared and specific representations for sequence $X$ are unified as:
\vspace{-1mm}
\begin{equation}
\begin{gathered}
    H_{\mathrm{sha}} = \sum_{n \leq R} \mathbf{q}_{t}^{M}[n]\, H_{n}^{M} + \sum_{n \leq R} \mathbf{q}_{t}^{A}[n]\, H_{n}^{A} + \sum_{n \leq R} \mathbf{q}_{t}^{B}[n]\, H_{n}^{B}, \\ H_{\mathrm{spec}}^{X} = \sum_{n>R} \mathbf{q}_{t}^{X}[n]\, H_{n}^{X},
\end{gathered}
\end{equation}
\vspace{-2mm}

\noindent where $\mathbf{q}_t[n]$ denotes the $n$-th element of the vector $\mathbf{q}_t$, and $X \in \{A, B, M\}$.

This strategy ensures that each single-domain sequence leverages both the shared experts and the most relevant domain-specific experts, effectively disentangling their contributions. For mixed sequences, all experts contribute, allowing comprehensive modeling across domains. Such an arrangement helps alleviate gradient conflicts and enhances representation learning for CDSR tasks.

\subsubsection{\textbf{Learning Objectives}}
To optimize the variational posterior distributions \(\mathbf{q}_t^A\), \(\mathbf{q}_t^B\), and \(\mathbf{q}_t^M\), a KL divergence regularization (\autoref{eq:elbo}) is employed. The learning objective is defined as:
\vspace{-1mm}
\begin{equation}
\mathcal{L}_{c} =  \sum_{X \in \{A, B, M\}} D_{\mathrm{KL}}\left( \mathbf{q}_t^X \| \frac{1}{V} \sum_{j=1}^V \mathbf{q}(S'^X_j) \right),
\end{equation}
\vspace{-1mm}

\noindent where \(\mathbf{q}_t^X\) represents the variational posterior for domain \(X \in \{A, B, M\}\), and \(\mathbf{q}(S'^X_j)\) denotes the produced variational posterior distribution with a pseudo sequence $S'^X_j$ as input, enforcing stability in dynamic expert assignment.

\subsection{\textbf{Variational Disentangling module}}
The variational disentangling module refines the preliminary separation achieved by the MoE encoders, targeting the mixed-sequence representation $H^M_{\mathrm{spec}}$ which may still contain entangled features. To further enhance the disentanglement of these components, we introduce two variational encoders, each implemented as a multilayer perceptron (MLP), which independently model the latent domain-specific preferences for each domain:
%
\begin{equation}
\mathbf{q}_{\phi_A}(z^A_{\mathrm{spec}}|H^M_{\mathrm{spec}}) = \mathcal{N}(\mu_A, \Sigma_A),\ \mathbf{q}_{\phi_B}(z^B_{\mathrm{spec}}|H^M_{\mathrm{spec}}) = \mathcal{N}(\mu_B, \Sigma_B),
\end{equation}
where $\mathcal{N}$ denotes normal distribution, $\mu$ and $\Sigma$ denote the mean and variance, respectively, $z^{\text{spec}}_A$ and $z^{\text{spec}}_B$ refer to the latent variables capturing the domain-specific preferences of domains $A$ and $B$. To enable differentiable sampling, we adopt the reparameterization trick:
\vspace{-2mm}
\begin{equation}
z^A_{spec} = \mu_A + \Sigma_A^{1/2} \cdot \epsilon_A, \quad
z^B_{spec} = \mu_B + \Sigma_B^{1/2} \cdot \epsilon_B,
\end{equation}
where $\epsilon_A,\epsilon_B$ are independent noise variables sampled from a standard normal distribution $\mathcal{N}(0, I)$. After obtaining the latent representations, we use a shared decoder $d_{\psi}(\cdot)$, also implemented as a MLP, to reconstruct the disentangled domain-specific components for domains $A$ and $B$:
\begin{equation}
H^{M \rightarrow A}_{spec} = d_{\psi}(z^A_{spec}), \quad
H^{M \rightarrow B}_{spec} = d_{\psi}(z^B_{spec}).
\end{equation}
The use of a shared decoder encourages the latent variables to capture domain-specific but structurally consistent information. 

\noindent\textbf{Learning Objectives.} To ensure effective disentanglement and informative latent representations, we also include KL divergence regularization terms in the overall loss function:
%
\begin{equation}
    \mathcal{L}_{\mathrm{var}} = \sum_{X=A,B} D_{KL}\big(\mathbf{q}_{\phi_X}(z^X_{\mathrm{spec}} | H^M_{\mathrm{spec}}) \| \mathcal{N}(0, I)\big).
\end{equation}

By regularizing these posteriors toward a prior \( \mathcal{N}(0, I) \), the model is guided to learn meaningful, structured, and disentangled domain-specific factors that can be leveraged for reconstructing domain-specific sequences.

\subsection{\textbf{Adversarial Disentangling Module}}
Although the Context-Aware MoE and the Variational Disentangling Module are introduced, relying solely on next-item prediction-oriented probability optimization methods (e.g., InfoNCE) remains insufficient to ensure complete disentanglement of representations and still risks negative transfer. Therefore, to enforce a more thorough and stable feature disentanglement, we incorporate an adversarial regularizer that leverages a domain discriminator coupled with a GRL. This regularizer performs top-down collaborative optimization of both the Variational Disentangling Module and the Context-Aware MoE. Specifically, we first fuse the domain-specific and domain-shared representations extracted from different sequences as follows:
\begin{equation}
\begin{gathered}
F_{sha} = f(H_{sha}), \\
F^A_{spec} = f(H^A_{spec} + H^{M \rightarrow A}_{spec}),\quad F^B_{spec} = f(H^B_{spec} + H^{M \rightarrow B}_{spec}),
\end{gathered}
\end{equation}
where $f(\cdot)$ denotes a non-linear transformation function (e.g., a feed-forward network with ReLU activation). The fused representations are then passed to a domain discriminator $D(\cdot)$, which is trained to distinguish their domain origin.

To train the domain discriminator, we assign ground-truth labels as follows: 
$F^{A}_{\text{spec}}$ is labeled as $0$ (domain A), 
$F^{B}_{\text{spec}}$ as $1$ (domain B), 
and $F_{\text{shared}}$ is assigned a soft label of $0.5$. To further prevent domain-specific information from leaking into the shared representation, we apply a GRL before feeding $F_{\text{shared}}$ into the discriminator. The GRL inverts the gradients during backpropagation, thereby confusing the discriminator and impeding its ability to accurately classify the domain of $F_{\text{shared}}$. This setting promotes a more complete disentanglement between shared and domain-specific features.

\noindent{\textbf{Learning Objectives.}}
The loss for this module is formulated as a domain classification loss:
\begin{equation}
\mathcal{L}_{\mathrm{adv}} = \mathcal{L}_{\mathrm{CE}}(D(\mathrm{GRL}(F_{sha})), y_{sha}) + \sum_{X=A,B} \mathcal{L}_{\mathrm{CE}}(D(F^X_{\mathrm{spec}}), y_X),
\end{equation}
where $\mathcal{L}_{CE}$ denotes the cross-entropy loss adapted for soft targets, $y_X \in\{0,1\}$ is the hard domain label for domain-specific features, and $y_{sha}$ is the soft label for the shared representation.

\subsection{Model Training}
\subsubsection{\textbf{Recommendation loss}} To train the model, we adopt the InfoNCE~\cite{oord2018representation} to compute the losses. Given the fusion representation $F$, its positive sample embedding $e^+$, and a set of $N_{\text{neg}}$ negative embeddings $E^-$ (sampled negative items from the same domain), the InfoNCE loss is defined as:
\begin{equation}
\text{InfoNCE}(F, e^+, E^-) = -\log \frac{\exp(F \cdot e^+ / \tau)}{\sum_{e \in \{e^+\} \cup E^-} \exp(F \cdot e / \tau)},
\end{equation}
where $\tau$ is a temperature hyperparameter controlling the softmax sharpness.

\noindent\textbf{Shared Preference Loss}.
To enhance the domain-invariance and representation quality of the shared preference vector $F_{\text{shared},i}$, we leverage its ability to predict the next item in the mixed-domain sequence $S^M$:
\vspace{-2mm}
\begin{equation}
\mathcal{L}_{\text{sha}} = \frac{1}{|S^M|} \sum_{i=1}^{|S^M|} \text{InfoNCE}\left(F_{sha,i}, e^+_{m,i}, E^-_{m,i}\right).
\end{equation}

\noindent\textbf{Domain-Specific Loss}.
For domain adaptation, we design a gradient-isolated joint prediction mechanism. Specifically, the domain-specific representation $F^A_{\text{spec},j}$ is combined with the gradient-stopped shared features $\text{SG}(F_\text{sha},j)$ to predict the next item in domain:
\begin{equation}
\mathcal{L}_{A} = \frac{1}{|S^A|} \sum_{j=1}^{|S^A|} \text{InfoNCE}\left(F^A_{\text{spec},j} + \text{SG}(F_{\text{sha},j}), e^+_{a,j}, E^-_{a,j}\right).
\end{equation}
The $B$-domain loss, $\mathcal{L}_{B}$, is computed analogously. The use of the stop-gradient operator $\text{SG}(\cdot)$ ensures that gradients from domain-specific losses $\mathcal{L}_{A}$ and $\mathcal{L}_{B}$ do not backpropagate into the shared representation. This design prevents conflicting supervision signals from different domains, thereby alleviating the gradient seesaw effect and stabilizing the optimization of domain-shared features.

\subsubsection{\textbf{Total Loss}}
The total loss combines recommendation losses and all regularization terms:
\begin{equation}
\mathcal{L}_{\text{total}} = \underbrace{\mathcal{L}_{\text{sha}} + \mathcal{L}_{A} + \mathcal{L}_{B}}_{\text{recommendation losses}} + \lambda_1 \mathcal{L}_{\text{c}} + \lambda_2 \mathcal{L}_{\text{var}} + \lambda_3 \mathcal{L}_{\text{adv}},
\end{equation}
where \(\lambda_1, \lambda_2, \lambda_3\) are hyperparameters that control the strength of the corresponding regularization terms. Their optimal values are reported in \autoref{sec: hyperparameter analysis}

\section{Experiments}
To comprehensively evaluate the effectiveness and robustness of CoDiS, we design a series of experiments to answer the following research questions:
\begin{enumerate}[leftmargin=8mm]
    \item[\textbf{RQ1}] How does CoDiS perform compared to SOTA baselines across different domains?
    \item[\textbf{RQ2}] What is the contribution of each key component in CoDiS?
    \item[\textbf{RQ3}] How generalized is CoDiS in cases of varying degrees of user overlap between domains?
    \item[\textbf{RQ4}] How robust is CoDiS against noise?
    \item[\textbf{RQ5}] How do critical hyperparameters influence CoDiS?
    \item[\textbf{RQ6}] How does CoDiS achieve effective disentanglement and causal backdoor adjustments in specific cases?
    \item[\textbf{RQ7}] How computationally efficient and scalable is CoDiS?
\end{enumerate}

\subsection{Experimental Setting}
The subsequent subsections detail the experimental setup, including datasets, baselines, evaluation protocols, and implementation details.
\subsubsection{\textbf{Datasets}}
Our experiments are conducted on three pairs of datasets in six distinct domains from the Amazon Product Reviews dataset\footnote{\url{https://cseweb.ucsd.edu/~jmcauley/datasets/amazon/links.html}}, containing user reviews and item metadata from May 1996
to July 2014. Specifically, the FK pair consists of "Grocery and Gourmet Food" as domain A and "Home and Kitchen" as domain B; the BE pair includes "Beauty" as domain A and "Electronics" as domain B; and the MB pair combines "Movies and TV" as domain A with "Books" as domain B.  ~\autoref{tab:dataset} summarizes the statistics of the datasets. 

For data pre-processing, each review is treated as a user interaction. We retain only users who have interacted with items in both domains of each domain pair, and remove items with fewer than five interactions to ensure data reliability. To further reduce computational overhead and mitigate cold-start noise, we truncate each user's interaction history to their 50 most recent actions, in line with prior work~\cite{kang2018self}. We employ the leave-one-out evaluation method to assess recommendation performance, consistent with prior studies~\cite{cao2022c2dsr}.

\vspace{-2mm}
\begin{table}[htbp]
\caption{CDSR Datasets Statistics.}
\vspace{-2mm}
\label{tab:dataset}
\centering
\resizebox{0.5\textwidth}{!}{%
\begin{tabular}{l|cc|cc|cc}
\toprule
\multirow{2}{*}{\textbf{Datasets}}
 & \multicolumn{2}{c|}{\textbf{FK}} &\multicolumn{2}{c|}{\textbf{BE}} & \multicolumn{2}{c}{\textbf{MB}} \\
 & Food(A) &Kitchen(B) & Beauty(A) &Electronics(B)&Movie(A) & 
Book(B) \\
\midrule
\textbf{Users} & \multicolumn{2}{c|}{7144} & \multicolumn{2}{c|}{4,474} & \multicolumn{2}{c}{28,350} \\
\textbf{Items} & 11,837 & 16,258 & 10,379 &  14,188 & 35,712 &  90,958 \\
\textbf{Interactions} & 83,663 & 89,885 & 50,329 &  63,800 & 347,654 & 403,147\\
\midrule
\textbf{Validation} & 2,837 & 4,307 & 2,086 & 2,388 & 11,728 & 16,622 \\
\textbf{Test} & 2,419 &4,725 &1,875 & 2,599 & 10,935 & 17,415 \\
\bottomrule
\end{tabular}
}
\end{table}
\vspace{-3mm}

\subsubsection{\textbf{Baselines}}
we compare CoDiS with four categories of baselines:

\noindent\textbf{ST-SDSR} (SASRec-1 and BERT4Rec-1): Single-task, single-domain sequential recommendation models, which are independently trained and evaluated on each domain.

\noindent\textbf{DT-SDSR} (SASRec-2 and BERT4Rec-2): Dual-task, single-domain sequential recommendation baselines, which are trained on both domains while maintaining domain-specific loss computation.
\begin{itemize}[leftmargin=1mm]
    \item \textbf{SASRec}~\cite{kang2018self}: a well-known SDSR baseline that employs self-attention to capture sequential patterns. We implement two variants: SASRec-1 and SASRec-2, where the latter computes and aggregates losses independently per domain to mitigate domain bias.
    \item \textbf{BERT4Rec}~\cite{Sun:2019:BSR:3357384.3357895}:  introduce Cloze objectives on top of SASRec. Similarly, we use two versions: BERT4Rec-1 and BERT4Rec-2.

\end{itemize}

\noindent\textbf{ST-CDSR} (CD-SASRec, CD-ASR~, and MGCL): Single-task, cross-domain sequential recommendation methods, which are trained on both domains, but only evaluated on the target domain. Thus, each model is executed twice, once targeting domain A and once targeting domain B.
\begin{itemize}[leftmargin=1mm]
\item \textbf{CD-SASRec}~\cite{alharbi2022cross}: a pioneering CDSR model that uses self-attention to learn source-domain representations and fuses them into target-domain sequence encoding.
    \item \textbf{CD-ASR}~\cite{alharbi2021cross}: combines source-domain multiplicative attention with target-domain self-attention to synthesize cross-domain sequential dynamics.
    \item \textbf{MGCL}~\cite{xu2025multi}: employs multi-view contrastive learning across graphical and sequential views from both cross-domain and domain-specific perspectives to enhance target-domain modeling.
\end{itemize}

\noindent\textbf{DT-CDSR} (C$^2$DSR, DREAM, and
ABXI): Dual-task, cross-domain sequential recommendation baselines, jointly trained and evaluated on both domains, with performance metrics separately computed per domain in a unified training process.
\begin{itemize}[leftmargin=1mm]
    \item \textbf{C$^2$DSR}~\cite{cao2022c2dsr}: uses separate graph and self-attention encoders to process cross-domain and domain-specific sequences, utilizing data augmentation techniques to facilitate contrastive learning.
    \item \textbf{DREAM}~\cite{2023dream}: constructs domain-specific sequential representations by extracting unique features per domain, then adaptively integrates them into the CDSR framework to better capture users’ cross-domain global preferences.
    \item \textbf{ABXI}~\cite{bian2025abxi}: employs a task-guided alignment strategy for domain-specific sequences and transfers domain-shared preference patterns into the modeling of domain-specific tasks.
\end{itemize}

\subsubsection{\textbf{Evaluation Protocol}}
 To ensure robustness, we evaluate all models using five random seeds with Hit Rate (HR) at K=5 and K=10, Normalized Discounted Cumulative Gain (NDCG) at K=10~\cite{jarvelin2002cumulated}, and Mean Reciprocal Rank (MRR)~\cite{voorhees-tice-2000-trec}. For single-target models, hyperparameters are optimized based on MRR within the respective domain. For dual-target models, hyperparameters are selected based on the aggregate MRR score across both domains. All hyperparameters of CoDiS are individually optimized for each dataset through comprehensive validation. The model is trained with a maximum of 600 epochs, including 10 warm-up epochs, and employs early stopping with a patience of 60 epochs. The AdamW optimizer is used throughout all experiments. Dimensionalities are specifically tuned according to the characteristics of each domain, while weight parameters and learning rates are independently adjusted to achieve optimal performance. The final configurations are determined based on the best validation results obtained for each specific dataset, which is detailed in \autoref{tab:hyperparameter}.
\vspace{-2mm}
\begin{table}[htbp]
\centering
\caption{Hyperparameter Selection}
\setlength{\abovecaptionskip}{-0.5mm}
\label{tab:hyperparameter}
\begin{tabular}{ll}
\toprule
\multicolumn{2}{l}{\textbf{Model Architecture}} \\
\hline
Embedding dim $h=256$ & Latent dim $d\in\{64,128\}$ \\
Encoder layers: 1 & Experts: $N=5$, $R=2$, $K=2$ \\
\hline
\multicolumn{2}{l}{\textbf{Loss Coefficients}} \\
\hline
$\lambda_1$ ($\mathcal{L}_c$): 0.3 & $\lambda_2$ ($\mathcal{L}_{\text{var}}$): 0.1 \\
$\lambda_3$ ($\mathcal{L}_{\text{adv}}$): 1.0 & Gradient reversal coeff.: 0.0--1.0 \\
\hline
\multicolumn{2}{l}{\textbf{Training Parameters}} \\
\hline
Temperature: & $\tau=0.75$ \\
Dropout rate: & 0.0--0.9 \\
Learning rate: & $\{5\times10^{-4}, 10^{-4}\}$ \\
Weight decay: & $\{5,2,1\}\times\{10^1,10^0,\ldots,10^{-3}\}, 0$ \\
LR decay: & $\times\{0.1\text{--}1.0\}$ after 30 stable epochs \\
\bottomrule
\end{tabular}
\vspace{-2mm}
\end{table}


\begin{table*}[!t]
\setlength{\abovecaptionskip}{0.5mm}
  \caption{Recommendation performance (RQ1): The highest scores are highlighted in bold, with the runner-up underlined. We assess statistical significance between CoDiS and the top baseline using paired t-tests (** for $p\leq0.01$).}
  \label{tab:overall performance}
\small
    \begin{tabular}{llcccccccc}
    \toprule
    \multirow{2}{*}{} & \multirow{2}{*}{\textbf{Methods}} & \multicolumn{4}{c}{\textbf{Beauty}} & \multicolumn{4}{c}{\textbf{Electronics}} \\
    \cmidrule(lr){3-6} \cmidrule(lr){7-10}
    & & HR@5 & HR@10 & NDCG@5 & MRR & HR@5 & HR@10 & NDCG@5 & MRR \\
    \midrule
    \multirow{2}{*}{\makecell[l]{ST-\\SDSR}} 
    & SASRec-1 & 0.1837$_{\pm0.0014}$ & 0.2597$_{\pm0.0038}$ & 0.1523$_{\pm0.0020}$ & 0.1295$_{\pm0.0016}$ & 0.1345$_{\pm0.0052}$ & 0.1894$_{\pm0.0038}$ & 0.1111$_{\pm0.0032}$ & 0.0982$_{\pm0.0033}$ \\
    & BERT4Rec-1 & 0.1687$_{\pm0.0043}$ & 0.2438$_{\pm0.0043}$ & 0.1404$_{\pm0.0034}$ & 0.1197$_{\pm0.0032}$ & 0.1277$_{\pm0.0067}$ & 0.1832$_{\pm0.0069}$ & 0.1053$_{\pm0.0054}$ & 0.0930$_{\pm0.0050}$ \\
    \midrule
    \multirow{3}{*}{\makecell[l]{ST-\\CDSR}} 
    & CD-SASRec & 0.1605$_{\pm0.0115}$ & 0.2530$_{\pm0.0166}$ & 0.1380$_{\pm0.0107}$ & 0.1162$_{\pm0.0085}$ & 0.1290$_{\pm0.0060}$ & 0.1842$_{\pm0.0030}$ & 0.1069$_{\pm0.0027}$ & 0.0948$_{\pm0.0032}$ \\
    & CD-ASR & 0.1661$_{\pm0.0065}$ & 0.2550$_{\pm0.0044}$ & 0.1424$_{\pm0.0033}$ & 0.1211$_{\pm0.0033}$ & 0.1355$_{\pm0.0049}$ & 0.1938$_{\pm0.0039}$ & 0.1122$_{\pm0.0033}$ & 0.0987$_{\pm0.0034}$ \\
    & MGCL & 0.1364$_{\pm0.0047}$ & 0.2109$_{\pm0.0078}$ & 0.1162$_{\pm0.0044}$ & 0.1001$_{\pm0.0035}$ & 0.1537$_{\pm0.0018}$ & 0.2159$_{\pm0.0042}$ & 0.1273$_{\pm0.0019}$ & 0.1118$_{\pm0.0012}$ \\
    \midrule
    \multirow{2}{*}{\makecell[l]{DT-\\SDSR}}
    & SASRec-2 & 0.2292$_{\pm0.0102}$ & 0.3262$_{\pm0.0032}$ & 0.1866$_{\pm0.0032}$ & 0.1530$_{\pm0.0032}$ & 0.1481$_{\pm0.0036}$ & 0.2169$_{\pm0.0041}$ & 0.1236$_{\pm0.0040}$ & 0.1058$_{\pm0.0039}$ \\
    & BERT4Rec-2 & 0.1970$_{\pm0.0027}$ & 0.3077$_{\pm0.0089}$ & 0.1679$_{\pm0.0057}$ & 0.1363$_{\pm0.0050}$ & 0.1519$_{\pm0.0049}$ & 0.2246$_{\pm0.0028}$ & 0.1275$_{\pm0.0031}$ & 0.1101$_{\pm0.0031}$ \\
    \midrule
    \multirow{4}{*}{\makecell[l]{DT-\\CDSR}} 
    & C$^2$DSR & 0.1835$_{\pm0.0066}$ & 0.2645$_{\pm0.0034}$ & 0.1519$_{\pm0.0038}$ & 0.1290$_{\pm0.0037}$ & 0.1288$_{\pm0.0072}$ & 0.1859$_{\pm0.0063}$ & 0.1081$_{\pm0.0047}$ & 0.0960$_{\pm0.0043}$ \\
    & DREAM & 0.2090$_{\pm0.0047}$ & 0.3043$_{\pm0.0069}$ & 0.1742$_{\pm0.0032}$ & 0.1447$_{\pm0.0032}$ & 0.1216$_{\pm0.0040}$ & 0.1817$_{\pm0.0041}$ & 0.1023$_{\pm0.0024}$ & 0.0895$_{\pm0.0019}$ \\
    & ABXI & \underline{0.2807}$_{\pm0.0082}$ & \underline{0.3835}$_{\pm0.0050}$ & \underline{0.2245}$_{\pm0.0043}$ & \underline{0.1846}$_{\pm0.0038}$ & \underline{0.1659}$_{\pm0.0021}$ & \underline{0.2389}$_{\pm0.0032}$ & \underline{0.1385}$_{\pm0.0014}$ & \underline{0.1200}$_{\pm0.0019}$ \\
    & \textbf{CoDiS} & \textbf{0.2929}$_{\pm0.0045}^{**}$ & \textbf{0.4030}$_{\pm0.0039}^{**}$ & \textbf{0.2341}$_{\pm0.0023}^{**}$ & \textbf{0.1901}$_{\pm0.0028}^{**}$ & \textbf{0.1872}$_{\pm0.0073}^{**}$ & \textbf{0.2653}$_{\pm0.0046}^{**}$ & \textbf{0.1546}$_{\pm0.0037}^{**}$ & \textbf{0.1327}$_{\pm0.0040}^{**}$ \\
    \midrule
    \midrule
    \multirow{2}{*} & \multirow{2}{*}{\textbf{Methods}} & \multicolumn{4}{c}{\textbf{Movie}} & \multicolumn{4}{c}{\textbf{Book}} \\
    \cmidrule(lr){3-6} \cmidrule(lr){7-10}
    & & HR@5 & HR@10 & NDCG@5 & MRR & HR@5 & HR@10 & NDCG@5 & MRR \\
    \midrule
    \multirow{2}{*}{\makecell[l]{ST-\\SDSR}}
    & SASRec-1 & 0.2258$_{\pm0.0031}$ & 0.2961$_{\pm0.0037}$ & 0.1647$_{\pm0.0025}$ & 0.1874$_{\pm0.0027}$ & 0.1357$_{\pm0.0029}$ & 0.1789$_{\pm0.0033}$ & 0.1007$_{\pm0.0022}$ & 0.1147$_{\pm0.0023}$ \\
    & BERT4Rec-1 & 0.2329$_{\pm0.0018}$ & 0.3105$_{\pm0.0012}$ & 0.1927$_{\pm0.0007}$ & 0.1696$_{\pm0.0027}$ & 0.1638$_{\pm0.0017}$ & 0.2152$_{\pm0.0013}$ & 0.1378$_{\pm0.0008}$ & 0.1243$_{\pm0.0006}$ \\
    \midrule
    \multirow{2}{*}{\makecell[l]{ST-\\CDSR}}
    & CD-SASRec & 0.2347$_{\pm0.0022}$ & 0.3117$_{\pm0.0026}$ & 0.1940$_{\pm0.0015}$ & 0.1709$_{\pm0.0017}$ & 0.1710$_{\pm0.0042}$ & 0.2253$_{\pm0.0033}$ & 0.1434$_{\pm0.0030}$ & 0.1285$_{\pm0.0026}$ \\
    & CD-ASR & 0.2352$_{\pm0.0045}$ & 0.3052$_{\pm0.0032}$ & 0.1956$_{\pm0.0027}$ & 0.1743$_{\pm0.0024}$ & 0.1622$_{\pm0.0010}$ & 0.2118$_{\pm0.0024}$ & 0.1372$_{\pm0.0010}$ & 0.1244$_{\pm0.0010}$ \\
    & MGCL & 0.2097$_{\pm0.0042}$ & 0.2851$_{\pm0.0040}$ & 0.1726$_{\pm0.0032}$ & 0.1509$_{\pm0.0033}$ & 0.1248$_{\pm0.0038}$ & 0.1668$_{\pm0.0049}$ & 0.1043$_{\pm0.0035}$ & 0.0946$_{\pm0.0033}$ \\
    \midrule
    \multirow{2}{*}{\makecell[l]{DT-\\SDSR}}
    & SASRec-2 & 0.2303$_{\pm0.0046}$ & 0.3067$_{\pm0.0043}$ & 0.1903$_{\pm0.0032}$ & 0.1673$_{\pm0.0030}$ & 0.1356$_{\pm0.0015}$ & 0.1830$_{\pm0.0014}$ & 0.1146$_{\pm0.0011}$ & 0.1034$_{\pm0.0010}$ \\
    & BERT4Rec-2 & 0.2317$_{\pm0.0008}$ & 0.3095$_{\pm0.0011}$ & 0.1925$_{\pm0.0015}$ & 0.1697$_{\pm0.0017}$ & 0.1547$_{\pm0.0014}$ & 0.2063$_{\pm0.0012}$ & 0.1302$_{\pm0.0009}$ & 0.1176$_{\pm0.0009}$ \\
    \midrule
    \multirow{4}{*}{\makecell[l]{DT-\\CDSR}}
    & C$^2$DSR & 0.2299$_{\pm0.0019}$ & 0.3003$_{\pm0.0026}$ & 0.1911$_{\pm0.0010}$ & 0.1700$_{\pm0.0005}$ & 0.1316$_{\pm0.0050}$ & 0.1767$_{\pm0.0050}$ & 0.1123$_{\pm0.0032}$ & 0.1025$_{\pm0.0028}$ \\
    & DREAM & 0.2507$_{\pm0.0068}$ & 0.3255$_{\pm0.0044}$ & 0.2082$_{\pm0.0040}$ & 0.1848$_{\pm0.0043}$ & 0.1469$_{\pm0.0037}$ & 0.1973$_{\pm0.0037}$ & 0.1237$_{\pm0.0033}$ & 0.1118$_{\pm0.0031}$ \\
    & ABXI & \underline{0.2859}$_{\pm0.0016}$ & \underline{0.3682}$_{\pm0.0030}$ & \underline{0.2388}$_{\pm0.0014}$ & \underline{0.2118}$_{\pm0.0011}$ & \underline{0.1973}$_{\pm0.0021}$ & \underline{0.2571}$_{\pm0.0019}$ & \underline{0.1669}$_{\pm0.0013}$ & \underline{0.1502}$_{\pm0.0012}$ \\
    & \textbf{CoDiS} & \textbf{0.2927}$_{\pm0.0012}^{**}$ & \textbf{0.3743}$_{\pm0.0022}^{**}$ & \textbf{0.2448}$_{\pm0.0009}^{**}$ & \textbf{0.2176}$_{\pm0.0012}^{**}$ & \textbf{0.2100}$_{\pm0.0013}^{**}$ & \textbf{0.2704}$_{\pm0.0012}^{**}$ & \textbf{0.1760}$_{\pm0.0011}^{**}$ & \textbf{0.1581}$_{\pm0.0011}^{**}$ \\
    \midrule
    \midrule
    \multirow{2}{*}{} & \multirow{2}{*}{\textbf{Methods}} & \multicolumn{4}{c}{\textbf{Food}} & \multicolumn{4}{c}{\textbf{Kitchen}} \\
    \cmidrule(lr){3-6} \cmidrule(lr){7-10}
     & & HR@5 & HR@10 & NDCG@5 & MRR & HR@5 & HR@10 & NDCG@5 & MRR \\
    \cmidrule(lr){1-10}
    \multirow{2}{*}{\makecell[l]{ST-\\SDSR}} & SASRec-1 & 0.1930$_{\pm0.0028}$ & 0.2611$_{\pm0.0036}$ & 0.1561$_{\pm0.0021}$ & 0.1332$_{\pm0.0019}$ & 0.1241$_{\pm0.0026}$ & 0.1851$_{\pm0.0018}$ & 0.1040$_{\pm0.0005}$ & 0.0900$_{\pm0.0007}$ \\
    & BERT4Rec-1 & 0.1819$_{\pm0.0035}$ & 0.2528$_{\pm0.0037}$ & 0.1462$_{\pm0.0027}$ & 0.1230$_{\pm0.0030}$ & 0.1114$_{\pm0.0040}$ & 0.1685$_{\pm0.0036}$ & 0.0926$_{\pm0.0029}$ & 0.0810$_{\pm0.0025}$ \\
    \cmidrule(lr){1-10}
    \multirow{3}{*}{\makecell[l]{ST-\\CDSR}} & CD-SASRec & 0.1797$_{\pm0.0079}$ & 0.2454$_{\pm0.0046}$ & 0.1421$_{\pm0.0060}$ & 0.1197$_{\pm0.0066}$ & 0.1119$_{\pm0.0067}$ & 0.1757$_{\pm0.0070}$ & 0.0946$_{\pm0.0045}$ & 0.0821$_{\pm0.0039}$ \\
    & CD-ASR & 0.1976$_{\pm0.0042}$ & 0.2727$_{\pm0.0052}$ & 0.1616$_{\pm0.0028}$ & 0.1368$_{\pm0.0026}$ & 0.1345$_{\pm0.0043}$ & 0.1995$_{\pm0.0044}$ & 0.1107$_{\pm0.0037}$ & 0.0941$_{\pm0.0034}$ \\
    & MCL & 0.1932$_{\pm0.0041}$ & 0.2673$_{\pm0.0054}$ & 0.1523$_{\pm0.0021}$ & 0.1260$_{\pm0.0018}$ & 0.1467$_{\pm0.0049}$ & 0.2157$_{\pm0.0026}$ & 0.1203$_{\pm0.0019}$ & 0.1017$_{\pm0.0019}$ \\
    \cmidrule(lr){1-10}
    \multirow{2}{*}{\makecell[l]{DT-\\SDSR}}& SASRec-2 & 0.2313$_{\pm0.0034}$ & 0.2854$_{\pm0.0049}$ & 0.1797$_{\pm0.0039}$ & 0.1535$_{\pm0.0034}$ & 0.1510$_{\pm0.0037}$ & 0.2168$_{\pm0.0049}$ & 0.1248$_{\pm0.0027}$ & 0.1062$_{\pm0.0021}$ \\
    & BERT4Rec-2 & 0.2223$_{\pm0.0053}$ & 0.2956$_{\pm0.0030}$ & 0.1727$_{\pm0.0033}$ & 0.1427$_{\pm0.0041}$ & 0.1363$_{\pm0.0043}$ & 0.2055$_{\pm0.0052}$ & 0.1116$_{\pm0.0032}$ & 0.0948$_{\pm0.0025}$ \\
    \cmidrule(lr){1-10}
    \multirow{4}{*}{\makecell[l]{DT-\\CDSR}}& C$^2$DSR & 0.1984$_{\pm0.0072}$ & 0.2574$_{\pm0.0116}$ & 0.1546$_{\pm0.0050}$ & 0.1311$_{\pm0.0035}$ & 0.1263$_{\pm0.0051}$ & 0.1879$_{\pm0.0061}$ & 0.1051$_{\pm0.0033}$ & 0.0903$_{\pm0.0027}$ \\
    & DREAM & 0.2158$_{\pm0.0043}$ & 0.2771$_{\pm0.0039}$ & 0.1698$_{\pm0.0025}$ & 0.1441$_{\pm0.0021}$ & 0.1377$_{\pm0.0021}$ & 0.2045$_{\pm0.0033}$ & 0.1138$_{\pm0.0012}$ & 0.0956$_{\pm0.0006}$ \\
    & ABXI & \underline{0.2498}$_{\pm0.0022}$ & \underline{0.3175}$_{\pm0.0041}$ & \underline{0.1973}$_{\pm0.0025}$ & \underline{0.1679}$_{\pm0.0028}$ & \underline{0.1737}$_{\pm0.0031}$ & \underline{0.2410}$_{\pm0.0026}$ & \underline{0.1415}$_{\pm0.0012}$ & \underline{0.1206}$_{\pm0.0014}$ \\
    & \textbf{CoDiS} & \textbf{0.2643}$_{\pm0.0035}^{**}$ & \textbf{0.3298}$_{\pm0.0025}^{**}$ & \textbf{0.2061}$_{\pm0.0026}^{**}$ & \textbf{0.1753}$_{\pm0.0025}^{**}$ & \textbf{0.1808}$_{\pm0.0027}^{**}$ & \textbf{0.2574}$_{\pm0.0031}^{**}$ & \textbf{0.1478}$_{\pm0.0027}^{**}$ & \textbf{0.1243}$_{\pm0.0031}^{**}$ \\
    \bottomrule
    \end{tabular}
\end{table*}
\begin{table*}[!t]
\setlength{\abovecaptionskip}{0.5mm}
\centering
\caption{Variants results on Food-Kitchen, Beauty-Electronics(RQ3).}
\label{tab:variants-results}
\renewcommand{\arraystretch}{1.2}
\resizebox{\linewidth}{!}{%
\begin{tabular}{l|ccc|ccc|ccc|ccc}
\toprule
\textbf{Variants} & \multicolumn{3}{c|}{\textbf{Food}} & \multicolumn{3}{c|}{\textbf{Kitchen}} & \multicolumn{3}{c|}{\textbf{Beauty}} & \multicolumn{3}{c}{\textbf{Electronics}} \\
 & MRR & NDCG@5 & HR@5 & MRR & NDCG@5 & HR@5 & MRR & NDCG@5 & HR@5 & MRR & NDCG@5 & HR@5 \\
\midrule
\textbf{CoDiS} & \textbf{0.1753}$_{\pm 0.0025}$ & \textbf{0.2061}$_{\pm 0.0026}$ & \textbf{0.2643}$_{\pm 0.0035}$  & \textbf{0.1243}$_{\pm 0.0031}$ & \textbf{0.1478}$_{\pm 0.0027}$ & \textbf{0.1808}$_{\pm 0.0027}$ & \textbf{0.1901}$_{\pm 0.0028}$ & \textbf{0.2341}$_{\pm 0.0023}$ & \textbf{0.2929}$_{\pm 0.0045 }$  & \textbf{0.1327}$_{\pm 0.0040 }$ & \textbf{0.1546}$_{\pm 0.0037}$ & \textbf{0.1872}$_{\pm 0.0073}$\\
w/o AL            & 0.1640 $_{\pm 0.0022  }$ & 0.1955$_{\pm 0.0021}$ & 0.2507$_{\pm 0.0011}$  & 0.1216$_{\pm 0.0017}$ & 0.1431$_{\pm 0.0019}$ & 0.1724$_{\pm 0.0024}$ & 0.1737$_{\pm 0.0041}$ & 0.2148$_{\pm 0.0042}$ & 0.2663$_{\pm 0.0078}$  & 0.1241 $_{\pm 0.0013}$ & 0.1444 $_{\pm 0.0013}$ & 0.1756 $_{\pm 0.0050}$ \\
w/o VD         & 0.1698 $_{\pm 0.0033}$ & 0.2002$_{\pm 0.0038}$ & 0.2562$_{\pm 0.0054}$  & 0.1209$_{\pm 0.0035}$ & 0.1434$_{\pm 0.0038}$ & 0.1745$_{\pm 0.0056}$ & 0.1836$_{\pm 0.0023}$ & 0.2243$_{\pm 0.0028}$ & \underline{0.2862}$_{\pm 0.0066}$  & 0.1198$_{\pm 0.0029}$ & 0.1398$_{\pm 0.0036}$ & 0.1726$_{\pm 0.0057}$ \\
w/o AL+VD         & \underline{0.1725}$_{\pm 0.0023}$  & \underline{0.2036}$_{\pm 0.0025}$  & 0.2570$_{\pm 0.0044}$ & 0.1223 $_{\pm 0.0017}$ & 0.1444 $_{\pm 0.0012}$ & 0.1776 $_{\pm 0.0018}$ & 0.1762 $_{\pm 0.0032}$ & 0.2179$_{\pm 0.0040}$ & 0.2703 $_{\pm 0.0070}$ & 0.1244 $_{\pm 0.0020}$ & 0.1436 $_{\pm 0.0018}$& 0.1731  $_{\pm 0.0022}$ \\
w/o CAR          & 0.1719 $_{\pm 0.0033  }$ & 0.2018$_{\pm 0.0032}$ & 0.2585$_{\pm 0.0032}$  & \underline{0.1233}$_{\pm 0.0022}$ & \underline{0.1463}$_{\pm 0.0022}$ & \underline{0.1796}$_{\pm 0.0029}$ & \underline{0.1850}$_{\pm 0.0037}$ & \underline{0.2272}$_{\pm 0.0041}$ & \underline{0.2862}$_{\pm 0.0071}$  & 0.1237$_{\pm 0.0018}$ & 0.1444$_{\pm 0.0017}$ & 0.1764$_{\pm 0.0056}$ \\
w/o EIS          & 0.1689 $_{\pm 0.0035  }$ & 0.2011$_{\pm 0.0031}$ & \underline{0.2592}$_{\pm 0.0016}$  & 0.1206$_{\pm 0.0031}$ & 0.1437$_{\pm 0.0037}$ & 0.1769$_{\pm 0.0061}$ & 0.1772$_{\pm 0.0084}$ & 0.2224$_{\pm 0.0091}$ & 0.2777$_{\pm 0.0185}$  & \underline{0.1255}$_{\pm 0.0034}$ & \underline{0.1470}$_{\pm 0.0042}$ & \underline{0.1788} $_{\pm 0.0073}$ \\
backbone         & 0.1687$_{\pm 0.0036  }$ & 0.1961$_{\pm 0.0034}$ & 0.2502$_{\pm 0.0020}$  & 0.1156$_{\pm 0.0015}$ & 0.1371$_{\pm 0.0012}$ & 0.1684$_{\pm 0.0018}$ & 0.1752$_{\pm 0.0031}$ & 0.2145$_{\pm 0.0022}$ & 0.2681$_{\pm 0.0036}$  & 0.1116$_{\pm 0.0020}$ & 0.1299 $_{\pm 0.0020}$ & 0.1567 $_{\pm 0.0017}$ \\
\bottomrule
\end{tabular}
}
\end{table*}

\vspace{-3mm}
\subsection{Overall Performance (RQ1)}

In this subsection, we analyze the performance of CoDiS by comparing it with various baseline methods. \autoref{tab:overall performance} shows a summary of the results, from which some key observations can be made.

First, across six domains from three datasets, CoDiS performs better than all baseline methods, including the latest CDSR models, on all metrics. Paired t-tests show that for all 24 metrics, the improvements are highly significant (p < 0.01).

Second, the relative improvements of CoDiS over baselines are more substantial in the B domains (kitchen, book, electronic), where the average percentage gain exceeds twice that of the A domains (beauty, movie, food). Importantly, A domains are characterized by richer data, while B domains tend to be much sparser. This result demonstrates that CoDiS effectively mitigates the "seesaw effect" and avoids negative transfer from data-rich to data-sparse domains.

Third, ST-CDSR models outperform ST-SDSR models on 19 out of 24 metrics, and DT-CDSR models also surpass DT-SDSR models, showing the effectiveness of the CDSR approach.

\vspace{-3mm}
\subsection{Ablation Study (RQ2)}
In this subsection, we conduct an ablation study to evaluate the necessity of the context-aware MoE encoders, the variational disentangled module, and the adversarial learning module, by progressively removing each component:
\begin{enumerate}[leftmargin=6mm]
    \item \textbf{w/o AL}: Remove adversarial learning module (-$\mathcal{L}_{adv}$)
    \item \textbf{w/o VD}: Remove variational disentanglement module (-$\mathcal{L}_{var}$)
    \item \textbf{w/o AL+VD}:
    Remove both adversarial learning module and variational disentanglement module
    \item \textbf{w/o CAR}: Replace context-aware router with uniform weights (-$\mathcal{L}_c$)
    \item \textbf{w/o EIS}: Disable expert isolation and selection mechanism
    \item \textbf{Backbone}: Pure MoE baseline without any proposed modules
\end{enumerate}

From \autoref{tab:variants-results}, we have several insightful observations: (1)All module variants outperform the backbone model but fall short of the full CoDiS framework, especially in sparse domains like kitchen and electronics. This confirms the collective role of our modules in transferring common preferences and reducing negative transfer. 
(2) Removing either adversarial learning (AL) or variational disentanglement (VD) alone causes a greater performance drop than removing both, indicating their synergistic effect for thorough disentanglement. (3) Replacing the context-aware router with uniform weights or disabling expert isolation/selection still beats the pure MOE backbone, yet underperforms compared to CoDiS. This outcome underscores the importance of context adjustment and expert isolation/selection in enabling each expert of the MOE to more accurately capture shared and specific preferences across varying contexts.

\subsection{Non-overlapping User Analysis (RQ3)}
\begin{figure}[htbp]
  \centering
\includegraphics[width=\linewidth]{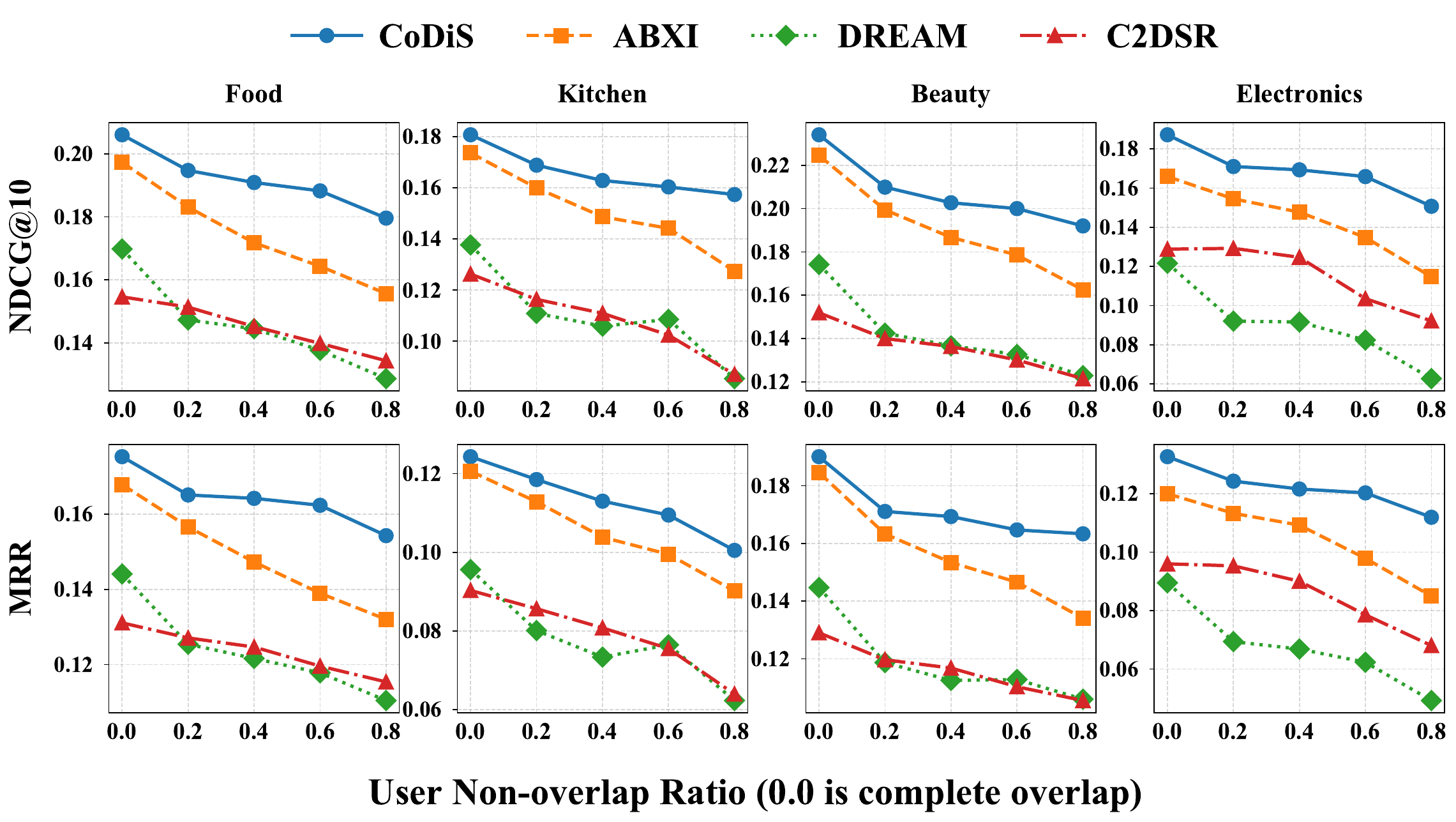}
  \caption{
    Performance of four CDSR models across domains under different user non-overlap ratios (RQ2).
  }
  \label{fig:non_overlap}
\end{figure}

In this subsection, we evaluate the robustness of CDSR models under varying user overlap ratios, implemented by selectively masking one domain's interactions for 0\% to 80\% of randomly chosen users. As shown in \autoref{fig:non_overlap}, CoDiS consistently outperforms all baselines and shows the slowest performance degradation as overlap decreases. These results confirm that CoDiS effectively extracts transferable cross-domain preferences through causal disentanglement, maintaining robust performance even with minimal user overlap.

\vspace{-2mm}
\vspace{-3mm}

\subsection{Causal Robustness Test (RQ4)}
Our robustness test, which injects 1–3 random items at random positions in interaction sequences, shows that CoDiS experiences significantly smaller performance degradation compared to ABXI across all domains (e.g., Beauty: 2.65\% vs. 8.06\%). This enhanced robustness stems from CoDiS’s ability to learn stable causal preferences—both domain-shared and domain-specific, while effectively eliminating confounding biases introduced by contextual noise.

\vspace{-1mm}
\begin{figure}[htbp]
    \centering
\includegraphics[width=\linewidth]{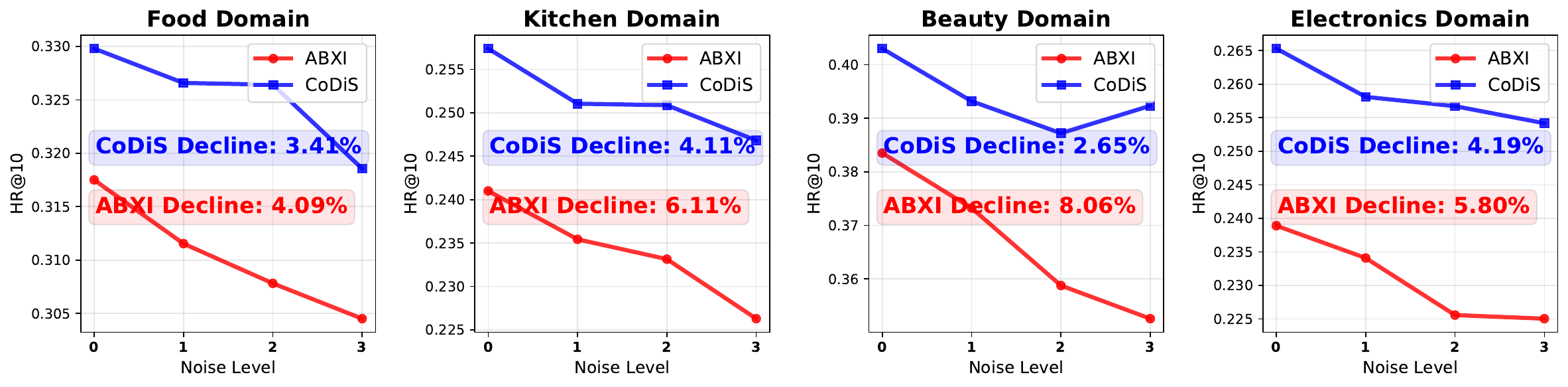}
    \caption{Performance Comparison under Increasing Noise.}
    \label{fig:noise}
\end{figure}
\vspace{-2mm}

\subsection{Hyperparameter Analysis (RQ5)}
\label{sec: hyperparameter analysis}
\noindent\textbf{Expert Configuration.}
The configuration of experts, specifically the total number of experts $N$, shared experts $R$, and selected specific experts $K$, significantly impacts the performance of CoDiS. 

\begin{figure}[htbp]
    \centering
\includegraphics[width=\linewidth]{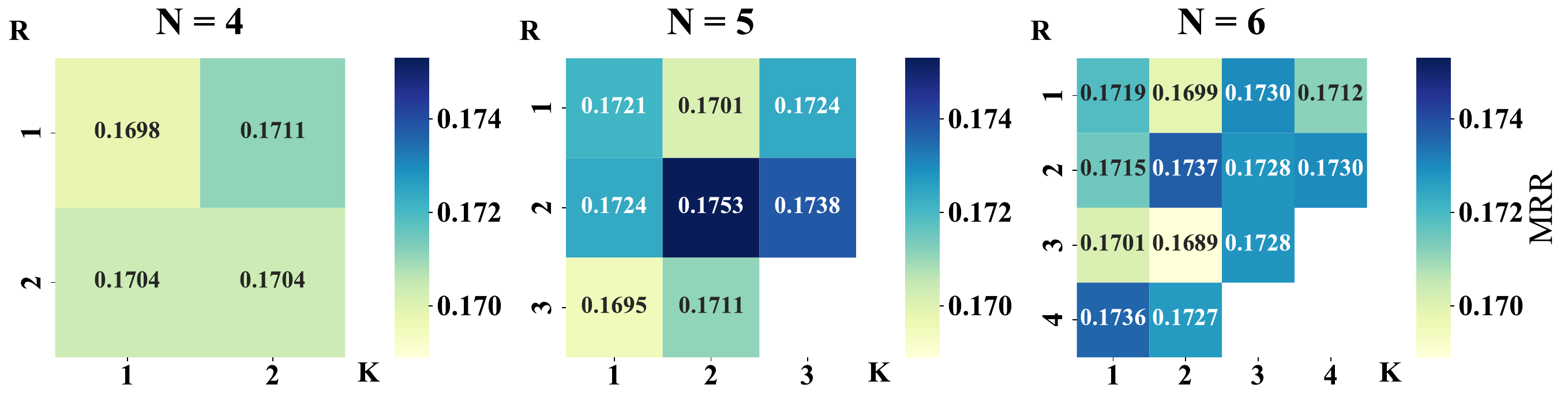}
    \caption{Impact of the number of total (N), shared (R), and specific (K) experts on kitchen MRR. Darker colors indicate better performance (RQ3).}
    \label{fig:hyparameters}
    \vspace{-2mm}
\end{figure}

As shown in ~\autoref{fig:hyparameters}, we observe the following: (1) Increasing $N$ allows the model to capture a richer variety of contexts, thereby enhancing its modeling capacity. However, the performance does not monotonically improve with $N$, indicating the need to balance between adequately modeling contextual information and avoiding overfitting. (2) The number of shared experts $R$ reflects the ability to capture general patterns across domains; too few shared experts may underfit, while too many could overshadow domain-specific patterns. (3) The best performance is achieved when $N$, $R$, and $K$ are proportionally balanced (e.g., $N=5$, $R=2$, $K=2$). This balance allows $N$ to capture diverse contexts without overfitting, while $R$ and $K$ together maintain an equilibrium between cross-domain sharing and domain-specific specialization.

\noindent\textbf{Regularization Trade-offs.}
To balance regularization terms in the total loss, the choice of \(\lambda_1\), \(\lambda_2\), and \(\lambda_3\) is crucial. The best performance is achieved at \(\lambda_1 = 0.3\), \(\lambda_2 = 0.1\), and \(\lambda_3 = 1.0\), and the performance generally follows a concave trend with respect to each \(\lambda\) (figure not shown due to space limits). This weighting achieves the desired trade-off between domain alignment and model specialization.

\subsection{Case Study (RQ6)}
\subsubsection{\textbf{Disentangled representation visualization}}
\begin{figure}[htbp]
  \centering
        \setlength{\abovecaptionskip}{-0.5mm}\includegraphics[width=\linewidth]{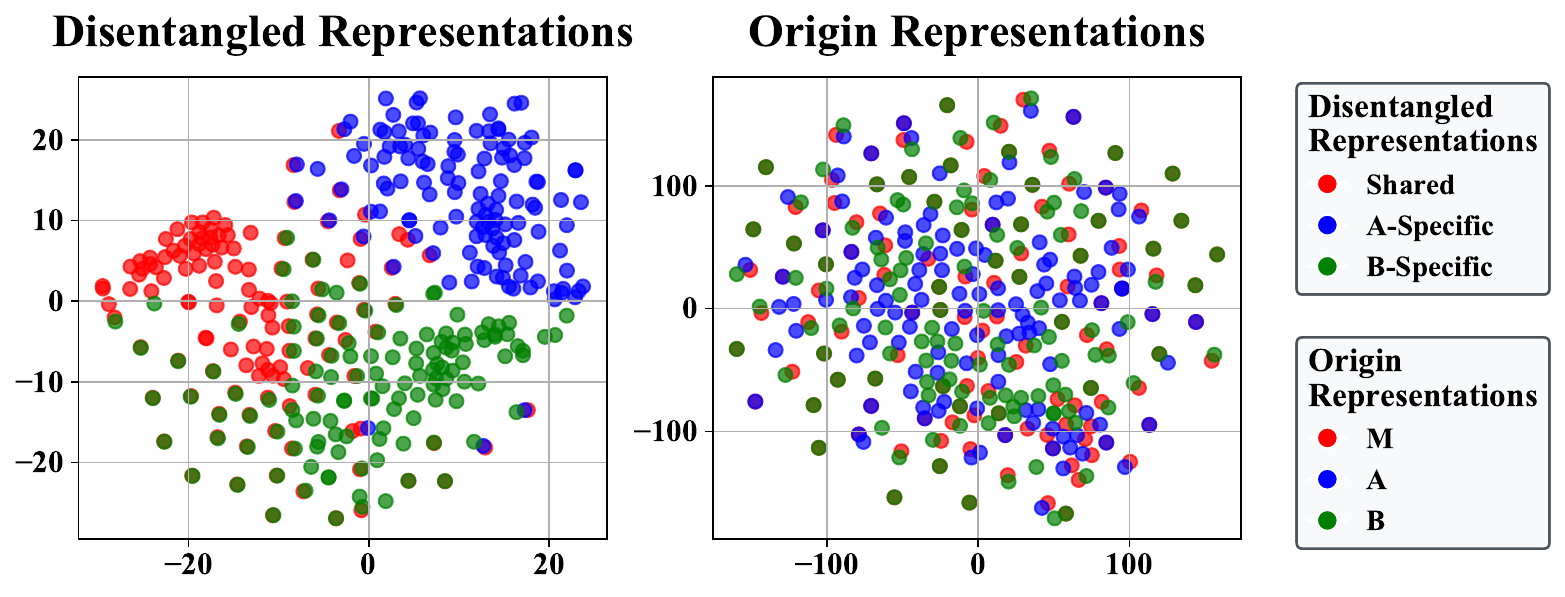}
  \caption{Comparison of users' disentangled and original representations (RQ5).}
\label{fig:comparison_representations}
\end{figure}

In this subsection, we randomly select several users and visualize their final disentangled domain-shared and domain-specific representations $F_{sha}, F^A_{spec}, F^B_{spec}$, alongside their original sequence embeddings $\mathbf{E}^M,\mathbf{E}^A,\mathbf{E}^B$, using t-SNE. The results (\autoref{fig:comparison_representations}) demonstrate that our model can effectively disentangle the domain-shared and domain-specific preferences, whereas the original sequence embeddings remain considerably entangled. 
\vspace{-2mm}
\vspace{-2mm}

\subsubsection{\textbf{Context Temproal Shift}}
\begin{figure}[htbp]
  \centering
  \includegraphics[width=\linewidth]{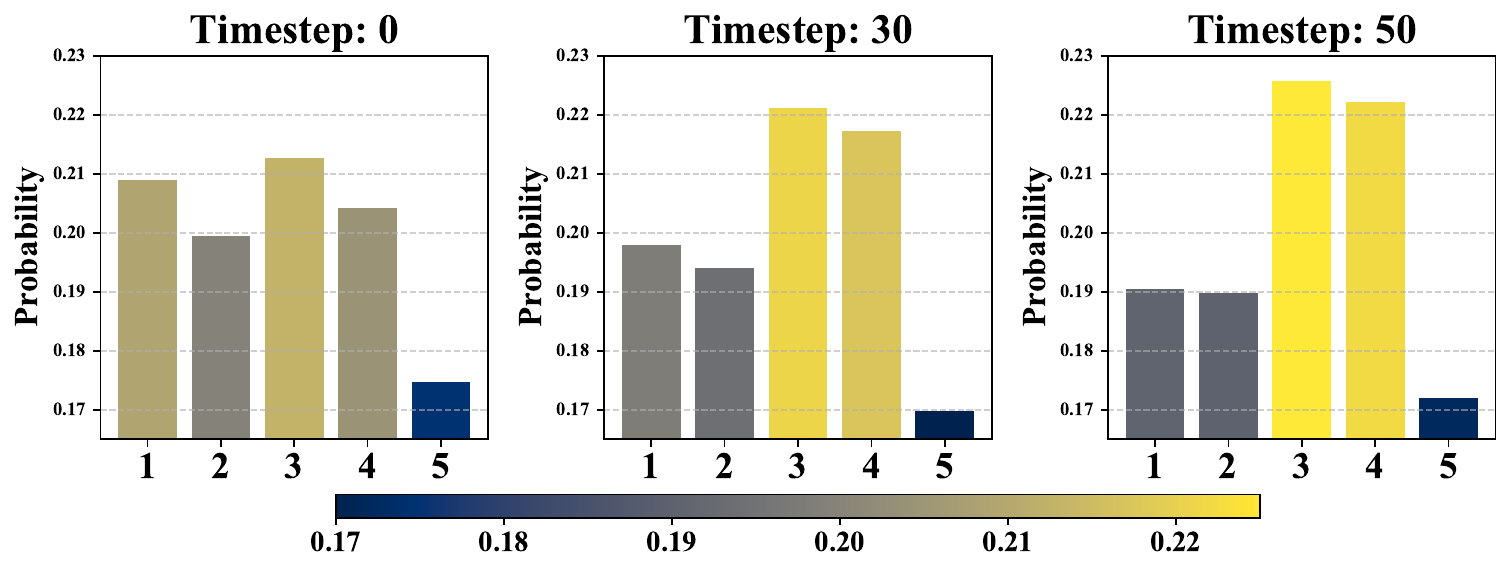}
  \caption{Visualization of probabilities for different contexts across timesteps (RQ5).}
\label{fig:prior_distribution}
\vspace{-3mm}
\end{figure}

To further investigate context distribution shift, in \autoref{fig:prior_distribution} we visualize the probabilities for each type of context at different time steps in a mixed sequence. The probability value for each context reflects its relative importance at a particular time. As shown in the figure, the probabilities of context 1 and context 2 gradually decrease over time, indicating that their importance diminishes as the sequence progresses. In contrast, the probabilities for context 3 and context 4 increase steadily, implying that these contexts become more dominant at later timesteps. These results clearly demonstrate the presence of temporal distribution shift across different time intervals. CoDiS first identifies the latent contexts underlying the data and effectively accounts for their shifting importance, thereby enabling more stable and robust prediction.
\vspace{-2mm}

\subsection{Time Complexity Analysis (RQ7)}
\label{sec: complexity}
A comparative analysis of computational efficiency per mini-batch (batch size: 256) is presented in ~\autoref{tab:time_complexity}.  CoDiS achieves SOTA performance, significantly outperforming the second-best model ABXI across key metrics, while maintaining highly reasonable computational costs. For the optimal configuration of 5 experts (N=5) identified on our primary dataset, CoDiS exhibits training time (0.316s) comparable to ABXI (0.288s), and near-identical inference time (0.0603s vs. 0.0607s). This indicates that for practical deployment, the superior performance of CoDiS comes at almost no additional inference-time cost.

Furthermore, the scalability of CoDiS is a key advantage. Even when confronting highly complex scenarios that require doubling the number of experts to 10 (also doubling the number of shared experts and selected experts), its training time (0.4012s) and inference time (0.0868s) remain within a practical range for real-world applications. This demonstrates that our model can dynamically adapt to increasing context demands without prohibitive computational overhead. Therefore, the minimal increase in computational cost is a justifiable and acceptable trade-off for the substantial performance improvement and enhanced scalability achieved by CoDiS.

\vspace{-2mm}
\begin{table}[htbp]
\centering
\caption{Time Complexity Analysis.}
\label{tab:time_complexity}
\resizebox{\linewidth}{!}{%
\begin{tabular}{l|cccc}
\toprule
\textbf{Metric} & \textbf{ABXI} & \textbf{DREAM} & \textbf{CoDiS(N=5)}& \textbf{CoDiS(N=10)} \\
\midrule
Training Time (s) & 0.288 & 0.225 & 0.316 &0.4012 \\
Inference Time (s) & 0.0607 & 0.0352 & 0.0603 & 0.0868 \\
\bottomrule
\end{tabular}
}
\end{table}

\section{Discussions and Conclusion}
\noindent
\textbf{Discussions}. The variational context adjustment framework approximates potentially discrete contexts (which may be infinitely many) with a fixed set to ensure tractability. 
This simplification is common among approaches that handle infinitely discrete confounders,
though our approach constitutes a step forward by scaling the context set to theoretically infinite. The computational overhead associated with this scaling is acceptable, as detailed in \autoref{sec: complexity}. Additionally, since the contextual variable is modeled as an abstract latent factor, assigning tangible, real-world semantics to it may be difficult or even impossible.

\noindent
\textbf{Conclusion}. This paper proposes CoDiS, a context-aware disentanglement framework for CDSR. CoDiS tackles challenges like contextual bias, gradient conflicts, and reliance on overlapping users. By combining context adjustment, expert selection, and variational adversarial disentanglement, it disentangles shared and domain-specific preferences to boost recommendation robustness. Experiments show that CoDiS outperforms existing methods and works well even with little user overlap while maintaining controllable computational complexity.

\bibliographystyle{ACM-Reference-Format}
\bibliography{mybib}









\end{document}